\documentclass[twocolumn,floatfix, showpacs]{revtex4}

\usepackage[final]{epsfig}
\usepackage{amsmath}
\usepackage{parskip}
 
\begin{document}
 
\title{Deflected Jets Can Not Explain the Double-Hump Structure in Triggered Correlations in Heavy-Ion Collisions }
 
\author{Thorsten Renk}
\email{thorsten.i.renk@jyu.fi}
\affiliation{Department of Physics, P.O. Box 35 FI-40014 University of Jyv\"askyl\"a, Finland}
\affiliation{Helsinki Institute of Physics, P.O. Box 64 FI-00014, University of Helsinki, Finland}
 

\begin{abstract}
Jet deflection by a flowing medium is one of the ideas brought forward to explain the splitting of the correlation function of hadrons associated with a high transverse momentum $P_T$ trigger from a jet-like structure observed in p-p and d-Au collisions to a double-hump structure on the away side in Au-Au collisions. However, just considering the parton kinematics  needed to explain the data shows that any attempt of detailed modelling deflected jets either cannot agree with the data or leads to internal contradictions and violations of basic physics principles. The idea that the deflection of jets by the medium can explain the experimentally observed structures should therefore be discarded.
\end{abstract}
 
\maketitle

Since first measurements of the correlation function of hadrons associated with a hard trigger in 200 AGeV Au-Au collisions at RHIC have shown a double-hump structure on the away side (i.e. opposite to the near side region defined by the trigger momentum vector) \cite{PHENIX-2pc}, very unlike the back-to-back jet peaks observed in p-p and d-Au collisions, the interpretation of this structure has been under intense debate. Early on, hydrodynamical shockwaves ('Mach cones') were among the likely candidates (see e.g. \cite{Stoecker, Solana}) as well as Cherenkov radiation \cite{Cherenkov}. The common feature of these mechanisms is that they are capable of producing the observed double-hump structure in each event, although it was later realized that the coupling of shockwaves to the collective flow of the medium is important and in a more realistic scenario including the effect of flow as well as the trigger bias on the hard vertex distribution not every shockwave leads to a double peak \cite{Mach1,Mach2,Mach3}. In contrast, deflected jets were suggested as a mechanism where in each event a single peak displaced from the away side direction $\phi=\pi$ is observed and the double-hump structure emerges only when averaging over many such events where the displacement occurs symmetrically around $\phi=\pi$ with both positive and negative deflection angles.

In order to avoid confusion of terminology, it is important to distinguish jet deflection as a theoretical concept that can in principle occur, jet deflection as an idea to explain the correlation measurements and, finally, jet deflection as a viable theoretical model to explain the observed correlations. 

As a theoretical concept, it has been argued in \cite{FlowJet} that the interaction of an in-medium parton shower with the surrounding medium should be sensitive to transverse flow, leading to the possibility of observing flow-distorted jet shapes. In \cite{Markovian} it has been shown that if one follows the trajectory of a single parton undergoing multiple scatterings with the medium, then there is a bias towards observing partons exiting the medium which have been deflected away from $\phi=\pi$ such that their in-medium pathlength is minimized, while partons scattered into the medium core are absorbed and hence cannot be observed.
 
Deflected jets as an idea to explain the observed correlation are more widespread and frequently encountered in presentations and experimental toy model simulations to be compared with data (see e.g. \cite{Ulery,Ajit}). However, there is no single realistic theoretical attempt to invoke deflected jets as the mechanism leading to the observed correlation structure which would use a medium description which is constrained by bulk observables or include a proper treatment of trigger bias. It is the purpose of this note to demonstrate that the reason for this lack of a realistic model is that such a model would need to violate basic physics principles in order to agree with the data or would lead to internal contradictions and that therefore also the idea of using deflected jets as a mechanism for creating a double-hump structure should be discarded.

Let us first consider the kinematics as constrained by the experimental observation. For simplicity, let us consider an event in which both back-to-back partons in the initial hard reaction are produced at midrapidity $\eta=0$ (the argument does not change when relaxing this assumption, only the notation becomes more complicated). Everything can then be described in the transverse $(x,y)$ plane where we take the $-x$ direction to be the direction of the trigger. A double-hump at large angles (albeit with an emerging central peak) is still seen for a trigger momentum range of 5-10 GeV in the associate momentum range of 2-3 GeV \cite{PHENIX-Detail}. When invoking jet deflection as an explanation, we must then assume that trigger and associate hadrons are produced in a fragmentation process (otherwise it is incorrect to speak of jets). The observed deflection angle is about 1.2 rad (other analyses find an even larger deflection angle of 1.36 rad \cite{Ulery}, but we consider the lower value in the following). Hadronization by parton fragmentation implies a probabilistic relation between parton and hadron momentum, but given the shapes of perturbative Quantum Chromodynamics (pQCD) parton spectrum and the fragmentation function, a mean fractional momentum $\langle z \rangle$ can be estimated as $\langle z \rangle \approx 0.7$.

The typical trigger hadron momentum then implies a fragmenting parton momentum of typically $p^{in}_{near}=(p_x, p_y) \sim (-9\, \text{GeV},0)$ and the away side parton momentum of $p^{in}_{away} \sim (9\,\text{GeV},0)$. In this estimate, effects like intrinsic $k_T$ (the momentum imbalance due to higher order pQCD effects and the initial state wave function) and intrinsic $j_T$ (the transverse difference between a parton momentum and the leading shower hadron momentum) have been neglected, as they average to a mean angular deflection of zero with an approximately Gaussian distribution around $\phi=0$, i.e. they do not contribute to explaining the double-hump structure, but just broaden any of the humps. The estimate also assumes that the typical near side parton energy loss is small. This is actually a result found in detailed energy loss calculations like \cite{EDep}

The finally observed away side hadron momentum on the other hand implies an away side parton momentum before fragmentation of $p^{f}_{away}\sim(1.6\,\text{GeV}, 3.2\,\text{GeV})$ (assuming in this case a deflection into the $+y$ direction). From this, the typical momentum change of the parton due to interactions with the medium can be estimated as $\Delta p_{med} \sim (-7.4\,\text{GeV}, 3.2\,\text{GeV})$, i.e. the medium dissipates most of the original momentum in $+x$ direction and transfers $\sim 3$ GeV momentum into $+y$ direction.

While it is entirely plausible and backed up by detailed calculations \cite{Back-to-Back} that a parton may lose $O(10)$ GeV of its momentum into the medium, it is quite implausible that a parton gains, on average, $O(3)$ GeV from the medium {\em into a given preferred direction}. Transverse flow identifies a preferred direction in the medium and could, combined with a geometrical bias, account for a deflection pattern away from $\phi=0$ --- but flow cannot accelerate a parton to a speed in excess of the flow velocity. The final away side parton has a velocity component $v_y = 0.93$ --- even assuming that the initial away side parton moves transverse to a flow field $(0, v_y)$ at all times, the typical flow velocities in the medium are not that large, especially at early times before hard partons escape from the medium.

One can also rephrase the statement: In order for a parton to gain on average $\Delta p_y = 3$ GeV from a medium where momenta are thermally distributed, the rest frame of the medium must move in such a way relative to the c.m. frame that the mean momentum of thermal partons seen in the c.m. frame is $\langle p_y \rangle \approx 3$ GeV. However, measured $P_T$ distributions of hadrons give no evidence for a mean parton $p_T$ of 3 GeV in the bulk medium.

Thus, a flowing medium cannot account for a deflection of a single parton trajectory strong enough to explain the data. This is to be contrasted with the effect of flow on a shockwave: Energy and momentum of a shockwave are distributed across many particles, each of which is in random motion. The shockwave itself is a property of the thermal medium not associated with any single parton, and hence propagates with the speed of sound relative to the thermal restframe --- thus is the thermal restframe is flow-boosted, the shockwave can be accelerated, even if no single parton is accelerated beyond the flow velocity.

One may think of invoking a different mechanism for the coupling between medium and jet to save the deflection picture, for example hadronization by recombination where a large part of the final hadron momentum is supplied by a parton picked up from the medium. However, any such scenario runs into a deeper conceptual problem: Why should a parton with a final momentum almost completely resulting from medium interactions be considered a deflected jet, rather than a part of the thermal medium?

To illustrate the situation clearly: If a parton can be almost stopped by the medium, and re-accelerated into a direction transverse to its original direction, and this parton can still be considered a jet, then any parton initially at rest in the medium can be accelerated to high $p_T$ into the deflection direction by the same mechanism and would also constitute a jet, as there is no way to distinguish a parton which has been stopped from one which has been at rest in the first place. The notion of jet deflection by a QCD medium is very different from the notion of, say, the deflection of a charged particle by a magnetic field. While in the latter case, the identity of the charged particle always remains distinct from the photon field, regardless of the deflection angle, a hard parton propagating through a partonic medium is distinct from the medium only through its momentum.

Note that also the measured correlation function does not tag the identity of the originally produced away side parton and and track the angular deflection of this tagged parton. Such a procedure may be viable in the case of heavy $c$ and $b$ quarks which are not thermally excited, but for light quarks, interactions with medium partons like $q\overline{q}\rightarrow gg$ may even change the identity of the leading jet parton without erasing the correlation information, because while parton identity is not conserved, momentum always is. Thus the experimental correlation function measures the flow of momentum and the redistribution of the original hard parton momentum by the interaction with the medium. While there is a bias for individual tagged parton trajectories to be deflected outward from the medium center as shown in \cite{Markovian}, this result has no strong connection with the measured correlation function. The measured correlation function is not sensitive to the trajectory of the original parton after many scatterings, it is sensitive to how the momentum lost from the original parton is redistributed. Thus, if a medium momentum transfer of $\Delta p_{med} \sim (-7.4\,\text{GeV}, 3.2\,\text{GeV})$ is needed to explain the data, it is evident that most of the original away side parton momentum is dissipated into the medium and that the redistribution mechanism of energy and momentum in the medium is responsible for the correlation and must be understood. That in itself is in manifest contradiction with the hypothesis that the correlation is created by the deflected original parton. 

Note that both the Cherenkov and the shockwave mechanism assume that the correlation is not created by the original parton but by the particular mechanism of energy and momentum redistribution. Thus, it is no problem for either of these scenarios that most of the measured final momentum comes from the medium, since the dispersion relations of the excitations determine the ratio of $p_x$ and $p_y$ in the medium recoil. In particular, note also that the momentum $P_y$ of the shockwave might be much less than the 3 GeV estimated for a deflected parton: In a deflected jet, all the energy and momentum of the finally observed hadron must come from a fragmenting parton, whereas a shockwave just boosts an already existing medium hadron to higher $P_T$, requiring significantly less energy.

There are yet more problems with a deflected jet scenario: Even if somehow the medium could cause such a strong deflection of a parton trajectory and the identity of the original parton would matter for the measured correlation, the data \cite{PHENIX-Detail} show that the deflection angle is remarkably unchanged as a function of both $P_T^{trigger}$ and $P_T^{assoc}$, even for trigger momentum range as low as 2-3 GeV. The variation in $\Delta p_{med}$ needed to keep the deflection angle fixed is substantial --- and yet $\Delta p_{med}$ is assumed to reflect properties of the medium like its temperature or flow velocity which do not change with trigger and associate momentum, leading to more conceptual problems. Finally, also note that the measured baryon/meson ratio in the correlation peak is incompatible with the assumption that the correlation is created by a deflected fragmenting parton \cite{McCumber}, although it may agree with a different hadronization mechanism for a hard parton.

In summary, both the distortion of jet shapes by flow and the bias for individual parton random-walk trajectories in the medium to bend away from the dense medium core are sound theoretical ideas. However, to invoke them as an explanation for the observed double-hump structure in the correlation implies the assumptions that flow can accelerate partons to velocities exceeding the flow velocity or that parton identity is conserved and matters for correlations rather than energy-momentum flow. Therefore jet deflection is not a possible mechanism even if toy models which do not take such complications into account may lead to correlations similar to those seen in the data.

\begin{acknowledgments}
 
 This work was supported by an Academy Research Fellowship from the Finnish Academy and from Academy Project 115262. 
 
\end{acknowledgments}

\end{document}